\documentclass[12pt]{article}
\newcommand{\sss}{\setcounter{equation}{0}}

\newtheorem{theorem}{THEOREM}[section]

\newtheorem{prop}[theorem]{PROPOSITION}

\def\ER{{\bf R}}
\def\beq{\begin{equation}}
\def\ene{\end{equation}}
\def\bull{\begin{flushright} \vrule height 6pt width 6pt depth -.pt
\end{flushright}}

\def\0{W_{1,2}^{(0)}}
\def\1{W_{-1,2}}
\def\2{W_{2,2}^{(0)}}
\def\u{e^{-itH}}

\begin{document}
%\baselineskip=21.6pt
% double space 28.8pt
%\baselineskip=28.8pt
\baselineskip=23.6pt
\title{The $L^{p}-L^{\acute{p}}$ Estimate for the Schr\"{o}dinger Equation on the Half-Line
 \thanks{2000 {\sc AMS} classification 35P25, 35R30  and 81U40. Research partially supported
by proyecto PAPIIT, IN 105799, DGAPA-UNAM.}}
\author{ Ricardo Weder \thanks{On leave of absence from Instituto de Investigaciones
en Matem\'aticas Aplicadas y en Sistemas, Universidad Nacional Aut\'onoma de M\'exico, 
Apartado Postal 20-726,
M\'exico D.F. 01000. Fellow, Sistema Nacional de Investigadores.}\\Instituto Argentino de 
Matem\'atica, CONICET\\
Saavedra 15, 1083-Buenos Aires, Argentina \\
E-Mail: weder@servidor.unam.mx\\}
\date{}
\maketitle

\begin{center}
\begin{minipage}{5.75in}
\centerline{{\bf Abstract}}\bigskip
In this paper we prove the $L^{p}-L^{\acute{p}}$ estimate for the  Schr\"{o}dinger 
equation on the half-line and with homogeneous Dirichlet boundary condition at the origin.

\end{minipage}
\end{center}

\newpage
\section{Introduction}\sss
We consider the following Schr\"{o}dinger equation,
\beq
i\frac{\partial}{\partial t}u(x,t)= H \,u(x,t),\,u(0,t)=0,\,
u(x,0)= \phi(x),
\label{1.1}
\ene
where $ x \in \ER^+:= (0, \infty),\, t \in \ER$. The Hamiltonian, $H$, is the following operator,
\beq
H:= -\frac{d^2}{dx^2}+ V(x),
\label{1.2}
\ene
with domain,
\beq
D\left(H \right):= \left\{ \phi \in L^2: \phi, \frac{d}{dx} \phi
\,\, \hbox{are absolutely continuos in}\, (0,\infty ),\, 
\left( -\frac{d^2}{dx^2}+ V(x)\right)\phi \in L^2,\phi(0)=0 \right\},
\label{1.3}
\ene
where $L^2$ denotes the Hilbert space of square-integrable functions on $\ER^+$. 
The potential, $V$, is real valued and it satisfies the 
following condition,

\beq
\int_0^{\infty} x \left|V(x)\right|\, dx < \infty.
\label{1.4}
\ene
The operator $H$ is self-adjoint in $L^2$ (see Section 2), it is the self-adjoint realization 
of the differential expression $ -\frac{d^2}{dx^2}+ V(x)$ with homogeneous Dirichlet boundary 
condition, $\phi(0)=0$, at zero. The unique solution to the initial-boundary value problem (\ref{1.1}) is 
given by,
\beq
u = e^{-itH} \phi,
\label{1.5}
\ene
where the strongly continuous one-parameter unitary group $e^{-itH}$ is defined by functional
calculus.  By $W_{l,p}, l=0,1,2,\cdots , 1 \leq p \leq \infty$, we denote the standard Sobolev
 spaces \cite{ad} in $\ER^+$ and by  $ W_{l,p}^{(0)}, 1 \leq p < \infty$, the completion of 
$C^{\infty}_0(\ER^+)$ in the 
norm of $ W_{l,p}$. The functions in $ W_{l,p}^{(0)}, l\geq 1$, satisfy the homogeneous 
Dirichlet boundary condition at zero, $\frac{d^j}{dx^j}u(0)=0, j=0,1,\cdots, l-1 $. In the case 
$l=0$ we use the standard notation, $W_{0,p}= W_{0,p}^{(0)}= L^p, p < \infty$.
 For $ l\geq 1$ we use the notation, $W_{l,\infty}^{(0)}:=
\{\phi \in W_{l,\infty}:
\frac{d^j}{d x^j}\phi(0)=0, j=0,1,2, \cdots, l-1\}$. 
Let us denote by $H_0$ the self-adjoint realization of $-\frac{d^2}{dx^2}$ with 
domain $ W_{2,2} \cap W_{1,2}^{(0)}$, i.e., the self-adjoint realization with homogeneous 
Dirichlet boundary condition at zero. It follows from a simple calculation using the Fourier transform that
$e^{-itH_0}$ is an integral operator,

\beq
e^{-itH_0}\phi = \int_0^{\infty}\, k_{t,0}(x,y)\, \phi (y)\, dy,
\label{1.6}
\ene
with the kernel,
\beq
k_{t,0}(x,y):=\frac{1}{2\pi}\int_{-\infty}^{\infty}\, e^{-itk^2}\,
\left[e^{ik(x-y)}-e^{ik(x+y)} \right]\, dk =
 \frac{1}{\sqrt{4\pi i t}}\, \left[e^{i(x-y)^2/4t}- e^{i(x+y)^2/4t}\right].
\label{1.7}
\ene
It follows from (\ref{1.6}) and (\ref{1.7}), integrating by parts, and as 
$\|\phi\|_{L^{\infty}}\leq \|\phi\|_{W_{1,1}}$,  that $e^{-itH_0}$ satisfies
the $L^1-L^{\infty}$ estimate,
\beq
\left\| e^{-itH_0}\right\|_{{\cal B}\left(W_{l,1}, W_{l,\infty}^{(0)}\right)}\leq C\,
\frac{1}{\sqrt{|t|}},
\,\, l=0,1,
\label{1.8}
\ene
and the $L^2-L^2$ estimate,
\beq
 \left\|e^{-itH_0} \right\|_{{\cal B}\left(W_{l,2}^{(0)}, W_{l,2}^{(0)}\right)}\
= 1, \,\, l=0,1,
\label{1.9}
\ene
where for any pair of Banach spaces, $X,Y$, we denote 
by ${\cal B}\left(X,Y\right)$ the Banach
space of bounded operators from $X$ into $Y$ and with $W_{0,\infty}^{(0)}:= L^{\infty}$.
 
Interpolating between (\ref{1.8}) and (\ref{1.9}) \cite{rs} we obtain the $L^p-L^{\acute{p}}$ estimate
in the free case, $V \equiv 0$,
\beq
 \left\|e^{-itH_0} \right\|_{{\cal B}\left(W_{l,p}^{(0)}, W_{l,\acute{p}}^{(0)}
\right)} \leq C \,\frac{1}{|t|^{(1/p-1/2)}},\,
 1/p+1/\acute{p}=1,\,1 \leq p \leq 2, \,   l=0,1.
\label{1.10}
\ene
It is clear that $e^{-itH}$ will  not   satisfy an estimate as (\ref{1.10}) if $H$ has 
eigenvectors. However, as we prove below it
satisfies the $L^p-L^{\acute{p}}$ estimate when restricted to the subspace of continuity of $H$,
${\cal H}_c$, i.e., to the subspace of $L^2$ orthogonal to all eigenvectors of $H$. We denote 
by $P_c$ the orthogonal projector onto ${\cal H}_c$. 

\begin{theorem}(The $L^p-L^{\acute{p}}$ estimate).
Suppose that $V$ satisfies (\ref{1.4}) then,
\beq
\left\| e^{-itH}\, P_c\right\|_{{\cal B}\left(L^p, L^{\acute{p}}\right)} \leq C \,\frac{1}{|t|^{(1/p-1/2)}}
,\, 1/p+1/\acute{p}=1,\,1 \leq p \leq 2.    
\label{1.11}
\ene
If, furthermore, $V\in L^1$,
\beq
\left\| e^{-itH}\, P_c\right\|_{{\cal B}\left(W_{1,p}^{(0)}, W_{1,\acute{p}}^{(0)}\right)}\leq C 
 \, \frac{1}{|t|^{(1/p-1/2)}},\, 1/p+1/\acute{p}=1,\,1 \leq p \leq 2.    
\label{1.12}
\ene
\end{theorem}   
We prove this result in Section 2 using the the Jost solution (the scattering solution) to the
stationary Schr\"{o}dinger equation. For the proof of the $L^p-L^{\acute{p}}$ estimate for the 
Schr\"odinger equation on the line see \cite{we2}, for the case of $\ER^n, n \geq 3$, see 
\cite{jss} and \cite{ya1}, \cite{ya2}. See \cite{ya3}, \cite{jy} for the problem in  $\ER^2$, with
 $1 < p \leq 2$. 
The $L^p-L^{\acute{p}}$ estimate expresses the {\it smoothing properties} of the linear 
Schr\"{o}dinger equation with a potential (\ref{1.1}) in a quantitative way, and it exibits the 
dispersive nature of this equation.
In fact, this estimate is of independent interest. As is well known, $L^p-L^{\acute{p}}$
estimates play an important role in the study of non-linear initial value problems
 \cite{st}, \cite{gi} and \cite{bou}. In particular,  the $L^p-L^{\acute{p}}$ estimate 
implies the famous Strichartz's 
estimates for the linear Schr\"{o}dinger equation with a potential, see \cite{str}, \cite{ka}, 
\cite{kt} and Proposition 2.4  below. 
It  is also the key  issue in scattering and inverse scattering  for non-linear Schr\"{o}dinger 
equations \cite{st}, \cite{gi}, \cite{bou}, \cite{ka}, \cite{jss}, \cite{we1}, \cite{we2}, \cite{we3}
and \cite{we4}. The $L^p-L^{\acute{p}}$ estimate is also important in the construction of
center manifolds for non-linear Schr\"odinger equations with potential, see \cite{sw}, 
\cite{pw} and \cite{we5}. In \cite{we6} we apply our $L^p-L^{\acute{p}}$ estimate to the solution of the 
direct and inverse scattering problems for the forced non-linear Schr\"odinger equation with a 
potential on the half-line.

\section{The $L^p-L^{\acute{p}}$\, Estimate}\sss
We first state a  number of results on the linear Schr\"odinger equation,
\beq
\left(- \frac{d^2}{dx^2}+ V(x)\right) \phi(x)= k^2 \,\phi(x), k \in{\mathbf C}.
\label{2.0}
\ene
Let us denote by $f(k,x), k \in {\mathbf C}, \hbox{Im} k \geq 0$, the Jost solution to (\ref{2.0}) (see
\cite{cs}, \cite{ma} and \cite{ne}). It is the solution to (\ref{2.0}) that satisfies
$f(k,x)\sim e^{ikx}, \,x \rightarrow \infty$. The Jost solution is not required to satisfy the homogeneous 
Dirichlet boundary condition at zero. It only satisfies it if $ k^2$ is an eigenvalue of $H$.  Let us denote, 
$\sigma (x):= \int_x^{\infty}|V(y)|\, dy $ and $ \sigma_1(x):= \int_x^{\infty}\sigma (y)\, dy$.
Condition (\ref{1.4}) is equivalent to $\sigma_1(0)< \infty$.   
The Jost solution can be represented as follows \cite{ma},

\beq
f(k,x)= e^{ikx}+ \int_x^{\infty} K(x,y )\, e^{ik y}\, dy,
\label{2.1}
\ene
where the kernel $K(x,y)$ is real valued and it satisfies the inequality,
\beq
\left|K(x,y)\right|\leq \frac{1}{2}\, \sigma\left(\frac{x+y}{2}\right)\, 
\exp\left(\sigma_1(x)-\sigma_1\left( \frac{x+y}{2} \right)\right),
\label{2.2}
\ene
and $K(x, y)=0, \, y < x$. 
Note that $\overline{f(k,x)}= f(-k,x), k \in \ER$.
For any $ k \in {\bf C}$ equation (\ref{2.0}) has also the regular solution, $\phi (k,x)$, that 
satisfies $\phi (k,0)=0, \, \acute{\phi}(k,0)=1$. Then, it follows from Theorem 5.8 of \cite{wei}
that $H$ is self-adjoint in the domain (\ref{1.3}). By the Parseval identity (see equation 
(3.2.4) in page 201 of \cite{ma}) we have that for all $ \phi \in L^2$,
\beq
 e^{-itH}\, P_c\,\phi = \int \,  k_t(x,y)\, \phi(y)\, dy,
 \label{2.3}
\ene
where,
\beq
  k_t(x,y):=\frac{1}{2\pi}\int_{-\infty}^{\infty}\, e^{-itk^2}\,\left[f(k,x)\, 
\overline{f(k,y)}
- f(k,x)\, f(k,y)\,S(k)\right]\, dk,
\label{2.4}
\ene
with $S(k)$ the "scattering matrix",
\beq
S(k):= \frac{f(-k,0)}{f(k,0)}.
\label{2.5}
\ene

We decompose $k_t(x,y)$ as follows,
\beq
k_t(x,y):= \sum_{j=0}^1  k_{t,j}(x,y),
\label{2.6}
\ene
where
\beq
k_{t,0}(x,y):=\frac{1}{2\pi}\int_{-\infty}^{\infty}\, e^{-itk^2}\,
\left[e^{ik(x-y)}-e^{ik(x+y)} \right]\, dk =
 \frac{1}{\sqrt{4 \pi i t}}\, \left[e^{i(x-y)^2/4t}- e^{i(x+y)^2/4t}\right],
\label{2.7}
\ene
corresponds to the free evolution with $V \equiv 0$ and, denoting, 
\beq
d(k,x):= f(k,x)-e^{ikx}, \,\,\, T(k):= S(k)-1,
\label{2.8}
\ene
we have that,
$$
k_{t,1}(x,y):=\frac{1}{2\pi}\int_{-\infty}^{\infty}\, e^{-itk^2}\,
\left[d(k,x)\, e^{-iky}+ e^{ikx}\, \overline{d(k,y)} +
d(k,x)\, \overline{d(k,y)}- d(k,x)\, e^{iky}-
\right.
$$
\beq
\left.
 e^{ikx}\, d(k,y) -
d(k,x)\,d(k,y)- T(k)\left( e^{ik(x+y)}+ d(k,x)\, e^{iky}+ e^{ikx}\, d(k,y) +
d(k,x)\,d(k,y)\right)\right]\, dk.
\label{2.9}
\ene
As indicated in the introduction (see (\ref{1.8})-(\ref{1.10})) the term with
$k_{t,0}(x,y)$ satisfies the the $L^p-L^{\acute{p}}$ estimate 
corresponding to the free case, $V\equiv 0$. We prove below a similar
statement for the term with $k_{t,1}(x,y)$.

\begin{theorem}(The $L^1-L^{\infty}$ estimate).
Suppose that $V$ satisfies (\ref{1.4}). Then,
\beq
\left\| e^{-itH}\, P_c\right\|_{{\cal B}\left(L^1, L^{\infty}\right)}\leq C
\frac{1}{\sqrt{|t|}}.
\label{2.10}
\ene
If moreover, $V \in L^1,$ then,
\beq
\left\| e^{-itH}\, P_c\right\|_{{\cal B}\left(W_{1,1}, W_{1,\infty}\right)}\leq C 
 \, \frac{1}{\sqrt{|t|}}.
\label{2.11}
\ene
\end{theorem}

\noindent{\it Proof:}\,\, We first prepare some results. Let us denote,
\beq
h(u,v):= K(u-v,u+v), u,v \geq 0, u \geq v.
\label{2.12}
\ene
Then,(see \cite{ma}, page 176), $h(u,v)$ is the unique solution to the following equation,
\beq
h(u,v)= \frac{1}{2}\int_u^{\infty}\, V(y)\, dy + \int_u^{\infty} \,dx\,  
\int_0^v \,V(x-y)\,h(x,y)\, dy.
\label{2.13}
\ene 
We denote,
\beq
q(u,v):=\frac{1}{2} \sigma(u)\, \exp\left(\sigma_1(u-v)-\sigma_1(u)\right), u \geq v. 
\label{2.14}
\ene 
We have that \cite{ma},
\beq
|h(u,v)| \leq q(u,v).
\label{2.14b}
\ene
Since  $q(u,v)$ is a non-increasing 
function of $u$, and as $q(u,v)$ is non-decreasing on $v$, we have that,
\beq
\left|\frac{\partial}{\partial u} h(u,v)\right| \leq \frac{1}{2} \left|V(u)\right|+  
\sigma(u-v)\, q(u,v),
\label{2.15}
\ene
and,
\beq
\left|\frac{\partial}{\partial v }\,h(u,v)\right| \leq \sigma(u-v)\, q(u,v).
\label{2.16}
\ene

We decompose $k_{t,1}$ as follows,
\beq
k_{t,1}(x,y)=k_{t,2}(x,y)+ k_{t,3}(x,y),
\label{2.16b}
\ene
where,
$$
k_{t,2}(x,y):=\frac{1}{2\pi}\int_{-\infty}^{\infty}\, e^{-itk^2}\,
\left[d(k,x)\, e^{-iky}+ e^{ikx}\, \overline{d(k,y)} +
d(k,x)\, \overline{d(k,y)}- d(k,x)\, e^{iky}-
\right.
$$
\beq
\left.
 e^{ikx}\, d(k,y) -
d(k,x)\,d(k,y)- T(k)\left( d(k,x)\, e^{iky}+ e^{ikx}\, d(k,y) +
d(k,x)\,d(k,y)\right)\right]\, dk.
\label{2.16c}
\ene
 and
\beq
k_{t,3}(x,y):= -\frac{1}{2\pi}\int_{-\infty}^{\infty}\, e^{-itk^2}\, T(k)\, e^{ik(x+y)}.
\label{2.16d}
\ene
Recall that,

\beq
f_t(z):= \frac{1}{2\pi}\int_{-\infty}^{\infty}\,e^{-itk^2}\,e^{-ikz}\, dz=
\frac{1}{\sqrt{4\pi i  t}}\, e^{i z^2/4t },
\label{2.20}
\ene
\beq
 \frac{1}{\sqrt{2\pi}}\int_{-\infty}^{\infty}\,d(k,x)\, e^{-ikz}\,dk= 
\sqrt{2\pi}\,K(x,z).
\label{2.21}
\ene
By (\ref{2.20}), (\ref{2.21}) and the convolution theorem for the Fourier transform,

\beq
b_t(x,y):= \frac{1}{2\pi}\int_{-\infty}^{\infty}\, e^{-itk^2}\,
d(k,x)\, e^{-iky} \, dk =\,\int_{x}^{\infty}\,f_t(y-z)\, K(x,z)\,dz.
\label{2.22}
\ene 
By (\ref{1.4}), (\ref{2.2}), (\ref{2.21}) and (\ref{2.22})

\beq
\left|b_t(x,y)\right|\leq \frac{C}{\sqrt{|t|}},
\label{2.23}
\ene
and moreover, if $V\in L^1$, it follows from
 (\ref{2.15}),(\ref{2.16}), (\ref{2.22}),
and since  $K(x,y)= h(\frac{x+y}{2}, \frac{y-x}{2})$ that,

\beq
\left|\frac{\partial }{\partial x}b_t(x,y)\right|\leq \frac{C}{\sqrt{|t|}}.
\label{2.24}
\ene
In the same way we prove that,
\beq
c_t(x,y):= \frac{1}{2\pi}\int_{-\infty}^{\infty}\, e^{-itk^2}\,e^{ikx}
\overline{d(k,y)}\, dk = \int_{y}^{\infty}\,f_t(x-z)\, K(y,z)\,dz,
\label{2.24b}
\ene 
and that,
\beq
\left|c_t(x,y)\right|\leq \frac{C}{\sqrt{|t|}}.
\label{2.24c}
\ene
Moreover, integrating by parts in $z$ we prove that,
\beq
\frac{\partial}{\partial x}c_t(x,y)= f_t(x-y)\, K(y,y)+ \int_{y}^{\infty}\,f_t(x-z)\, 
\frac{\partial}{\partial z}K(y,z)\,dz,
\label{2.24d}
\ene 
and it follows that,

\beq
\left|\frac{\partial }{\partial x}c_t(x,y)\right|\leq \frac{C}{\sqrt{|t|}}.
\label{2.24e}
\ene

Similarly,
\beq
e_t(x,y):= \frac{1}{2\pi}\int_{-\infty}^{\infty}\, e^{-itk^2}\,
d(k,x)\, \overline{d(k,y)}\, dk = \sqrt{2\pi}\,\int \,f_t(z_1)\, K(x,z_1-z_2) 
\, K(y,-z_2)  \,dz_1
\, dz_2,
\label{2.25}
\ene 
and we prove as above that if (\ref{1.4}) is satisfied then,

\beq
\left|e_t(x,y)\right|\leq \frac{C}{\sqrt{|t|}},
\label{2.26}
\ene
and, if $V \in L^1$, 
\beq
\left|\frac{\partial }{\partial x}e_t(x,y)\right|\leq \frac{C}{\sqrt{|t|}}.
\label{2.27}
\ene
We estimate the remaining terms in the right hand side of (\ref{2.16c}) in the same way. For 
this 
purpose, note that it follows from the extension of  Wiener's theorem to the Fourier transform
(see the Corollary to Theorem 4.204 in page 154 of \cite{hp} ) and the argument given in 
pages 212 and 213 of \cite{ma} that the Fourier transform of $T(k)$ is integrable on 
$(-\infty,\infty)$. Then, if (\ref{1.4}) holds,
\beq
\left|k_{t,2}(x,y)\right|\leq \frac{C}{\sqrt{|t|}},
\label{2.28}
\ene
and, if $V \in L^1$, 

\beq
\left|\frac{\partial }{\partial x}k_{t,2}(x,y)\right|\leq \frac{C}{\sqrt{|t|}}.
\label{2.29}
\ene
Furthermore, let us denote by $T_{t,3}$ the integral operator with kernel $k_{t,3}(x,y)$.
Then denoting by $\tilde{T}$ the inverse Fourier transform of $T$, we have that, 
 
\beq
T_{t,3}\phi = \int_{-\infty}^{\infty}\,dz\,\tilde{T}(z) \,\int_0^{\infty} f_t(x+y-z) 
\phi(y)\, dy.
\label{2.29b}
\ene
Hence,
\beq
\left\| T_{t,3} \right\|_{\mathcal{B}(L^1, L^{\infty})}\leq C \frac{1}{\sqrt{|t|}},
\label{2.29c}
\ene
and integrating by parts in $y$ and as $\|\phi\|_{L^{\infty}}\leq \|\phi\|_{W_{1,1}}$, 
 we prove that,

\beq
\left\| T_{t,3} \right\|_{\mathcal{B}(W_{1,1}, W_{1,\infty})}\leq C \frac{1}{\sqrt{|t|}}.
\label{2.29d}
\ene
The Theorem follows from (\ref{1.8}), (\ref{2.3}),(\ref{2.6}), (\ref{2.16b}), (\ref{2.28}), 
(\ref{2.29}), (\ref{2.29c}) and (\ref{2.29d}).

\begin{prop}
Suppose that
\beq
\sup_{x \in \ER}\int_x^{x+1}|V(y)|\, dy < \infty.
\label{3.34b}
\ene

Then, for any $\epsilon >0$ there is a constant, $K_{\epsilon}$, such that,
\beq
\int_0^{\infty} |V(x)|\,  |\phi(x)|^2\, dx \leq \epsilon \left\|\acute {\phi} \,\right\|_{L^2}^2+ K_{\epsilon} \left\|\phi \right\|_{L^2}^2, \phi \in W_{1,2}.
\label{3.34c}
\ene
\end{prop}

\noindent{\it Proof:} \,\,  If  $\phi \in W_{1,2 }$,  for any $n=0,1, \cdots$,
 any $x,y \in [n, n+1]$ and any $\delta > 0$, we have that,
\beq
|\phi(x)|^2- |\phi(y)|^2 = 2  \, \hbox{Re} \int_y^x \phi(z) \overline{\acute{\phi}(z)}\, dz
\leq \delta \int_n^{n+1}|\acute{\phi(z)}|^2 dz + \frac{1}{\delta} \int_n^{n+1}|\phi(z)|^2 dz.
\label{3.35c}
\ene
By the mean value theorem we can choose $y$ such that,  $|\phi(y)|^2= \int_n^{n+1}|\phi(z)|^2 dz$, and it follows that,
\beq
|\phi(x)|^2 
\leq \delta \int_n^{n+1}|\acute{\phi(z)}|^2 dz + \left(1+ \frac{1}{\delta}\right) \int_n^{n+1}|\phi(z)|^2 dz.
\label{3.35d}
\ene
Let $C$ be the finite quantity in the left-hand side of (\ref{3.34b}). Then,
\beq
\int_n^{n+1} |V(x)| |\phi(x)|^2\, dx \leq  C \delta  \int_n^{n+1}|\acute{\phi(z)}|^2 dz + C \left(1+ \frac{1}{\delta}\right) \int_n^{n+1}|\phi(z)|^2 dz.
\label{3.35e}
\ene
Taking $\delta$ so small that  $  \epsilon= \delta C$, and adding over $n$  we obtain (\ref{3.34c}).

\bull

If $V$ satisfies (\ref{3.34b}) -in particular if $V \in L^1$-  it follows from (\ref{3.34c}) that the quadratic form, $ {\cal H}(\phi, \psi):=(\acute{\phi}, \acute{\psi})_{L^2}
+(V \phi, \psi)_{L^2}$ with domain $\0$ is closed and bounded below. Furthermore, $H$ is 
the associated self-adjoint operator (cf Theorem X.17 of \cite{rs}). Hence, the form domain 
of $H$ is $\0$,
and, $D\left( \sqrt{H+N}\right)= \0$, where $N$ is so large that, $H+N > 0$.
By (\ref{3.34c}) and as $ \left( H+N  \right)^{-1/2}$ is bounded from $L^2$ into $\0$
the norm $\left\|\sqrt{H+N}\,\phi \right\|_{L^2}$ is equivalent to the norm of $\0$. We use this
equivalence below without further comment. We denote by $P_{pp}$ the projector onto the pure 
point subspace of $H$.

\begin{theorem}(The $L^2-L^{2}$ estimate).
Suppose that $V$ satisfies (\ref{1.4}) and let us denote, $P_1:= I$, $P_2:= P_{pp}$ and
$P_3:= P_c$ . Then,
\beq
\left\| e^{-itH}\, P_n\right\|_{{\cal B}\left(L^2, L^2\right)}\leq C \,, n=1,2,3.
\label{2.30}
\ene
If furthermore, $V\in L^1$,

\beq
\left\| e^{-itH}\, P_n\right\|_{{\cal B}\left(\0, \0\right)}\leq C.
\label{2.30b}
\ene

\end{theorem}

\noindent {\it Proof:}\, Equation (\ref{2.30}) is just the unitarity of $e^{-itH}$ in $L^2$.
Equation (\ref{2.30b}) with $n=1$  follows from $ \sqrt{H+M}\u = \u \sqrt{H+M}$.
 It is proven in \cite{ma} that $H$ has a finite number 
of eigenvalues, $  E_1, E_2, \cdots, E_Q$, with $ E_j= (i k_j)^2$, and where $ 0 < k_1 < k_2, 
\cdots, k_ Q$ are the zeros of $f(k,0)$ i.e., $ f(k_j,0)=0, 1 \leq j \leq M$. We use the notation,
$\hat{f}_j(x):= f(ik_j,x)/ \|f(ik_j,\cdot)\|_{L^2}$. Then, the projector onto the pure point 
subspace of $H,\, P_{pp}$, is given by, 
\beq
 P_{pp}\, \phi= \sum_{j=1}^Q \hat{f}_j(x)\, (\phi,\hat{f}_j ).
\label{2.33}
\ene
As $\hat{f}_j \in \0$, we have that 
$P_{pp} \in \mathcal{B}(L^2,\0)$. Hence, (\ref{2.30b}) with $n=2$ holds because $P_{pp}$
commutes with $\u$. The case $n=3$ follows since $P_c= I-P_{pp}$.

\noindent {\it Proof of Theorem 1.1:}  Equation (\ref{1.11}) 
follows from (\ref{2.10}), (\ref{2.30}) with $n=3$ and interpolation \cite{rs}.
By (\ref{2.11}), (\ref{2.30b}) with $n=3$ and approximating
$\phi \in W_{1,1}^{(0)}$ by a sequence $\phi_n \in W_{1,1}^{(0)} \cap \0$ we prove that
 \beq
\left\| e^{-itH}\, P_c\right\|_{{\mathcal B}\left(W_{1,1}^{(0)},W_{1,\infty}^{(0)}\right)} \leq C 
 \, \frac{1}{\sqrt{|t|}}.
\label{2.45b}
\ene
By (\ref{2.30b}) with $n=3$, (\ref{2.45b}) and interpolation, (\ref{1.12}) holds. 
\bull

The $L^p-L^{\acute{p}}$ estimate implies that $e^{-it H}\, P_c$ has smoothing effects that are 
expressed by its action between certain function spaces. Let us denote,

\beq
L^{p,r}:= L^r\left(I, L^p\right),\, 1 \leq r, p \leq \infty, \, I:=[0,T], 
T > 0.
\label{2.46}
\ene   

In the case $T = \infty$ we take $I:=[0, \infty)$. We find it convenient to represent the pair 
$p,\, r$ by the point $P:=(1/p,1/r )$ in the square,
$[0,1]\times [0,1]$. Let us denote, $L(P):= L^{p,r}$. 
Let $s$ be the closed segment connecting $B:=(1/2, 0)$ and 
$C:=(0, 1/4)$. The equation for $s$ is: $ 1/p+2/r = 1/2,
0 \leq 1/p \leq 1/2$. For any $P= (1/p,1/r)\in s $ the dual point $\acute{P}$ is defined as
$\acute{P}:= (1/\acute{p}, 1/\acute{r})$ where, $1/p+ 1/ \acute{p}=1, 1/r+ 1/\acute{r}=1$.
$\acute{P}$ is on the dual segment, $\acute{s}$, connecting $\acute{B}=(1/2,1)$ to $\acute{C}=
(1,3/4)$. 
Let us define the following linear operators,
\beq
\left( \Gamma \phi \right)(t):= e^{-it H}\, P_c \,\phi, t \in I,
\label{2.47}
\ene

and
\beq
\left( \mathcal{G}f \right)(t):= \int_0^{t}\, e^{-i(t-\tau)H} \, P_c\, f(\tau)\, d\tau,
 \, t \in I.
\label{2.48}
\ene   

\begin{prop}(Strichartz's estimate)
Suppose that (\ref{1.4}) holds. Then $\Gamma $ is a bounded operator from $L^2$ into every
$L(P), P \in s$ and $\Gamma^{\ast}$ is bounded from every $L(Q), Q \in \acute{s}$ into $L^2$
with operator norm independent of $T$. Moreover, 
$\mathcal{G}$ is bounded from any $L(Q), Q \in \acute{s}$
to any $L(P), P \in s$ with operator norm  
independent of $T$.
\end{prop}

\noindent{\it Proof:}\, the proposition follows from (\ref{1.11}) and Theorem 1.2 of \cite{kt}.
\bull
\noindent Note that in Proposition 2.4 we can replace $L(B)$ by the space, 
$\overline{L}(B)$, of the bounded and
continuous functions from $I$ into $L^2$.


\begin{thebibliography}{99}
\bibitem{ad}R.A. Adams, "Sobolev Spaces",Academic Press, New York, 1970.

\bibitem{bou} J. Bourgain, "Global Solutions of Nonlinear Schr\"odinger Equations",
Colloquium Publications {\bf 46}, Amer. Math. Soc., Providence, RI, 1999.

 
\bibitem{cs} K. Chadan and P.C. Sabatier, "Inverse Problems in Quantum Scattering Theory. Second 
Edition", Springer, Berlin,1989.



%\bibitem{fr} A. Friedman, "Partial Differential Equations", Holt, Rinehart and Winston Inc., New York, 1969. 

\bibitem{gi} J. Ginibre, " Introduction aux \'Equations de Schr\"odinger nonlin\'eares",
Onze Editions, Paris, 1998.

\bibitem{hp} E. Hille and R.S. Phillips, "Functional Analysis and Semigroups", Colloquium
Publications {\bf XXI}, Amer. Math. Soc., Providence, 1957.

\bibitem{jy} A. Jensen and K. Yajima, A remark on $L^p$-boundedness of wave operators for 
two dimensional Schr\"{o}dinger operators, Comm. Math. Phys. {\bf 225} (2002), 633-637.

\bibitem{jss}J.L. Journ\'e, A. Soffer and C.D. Sogge, Decay estimates for  Schr\"odinger 
operators, Comm. Pure Appl. Math. {\bf 44} (1991), 573-604.  

\bibitem{ka} T. Kato, Nonlinear Schr\"odinger equations, Lecture Notes in Physics {\bf 345},
(1989), 218-263, Springer, Berlin. 

\bibitem{kt}M. Keel, and T. Tao, Endpoint Strichartz estimates, Amer. J. Math. {\bf 120}
(1998), 955-980.

 \bibitem{ma} V.A. Marchenko, "Sturm-Liouville Operators and Applications", Birkh\"auser, 
Basel, 1986.



\bibitem{ne} R.G. Newton, "Scattering Theory of Waves and Particles. Second Edition", 
Springer, Berlin, 1982.

\bibitem{pw} C.-A. Pillet and C.E. Wayne, Invariant manifolds for a class of dispersive, 
hamiltonian, partial differential equations, J. Differential Equations {\bf 141} (1997),
310-326.

\bibitem{rs} M. Reed and B. Simon, "Methods of Modern Mathematical Physics II Fourier Analysis,
Self-Adjointness", Academic Press, New York, 1975.


\bibitem{st} W.A. Strauss, "Nonlinear Wave Equations", CBMS-RCSM {\bf 73}, Amer. Math.
Soc., Providence, R.I., 1989.

\bibitem{str} R.S. Strichartz, Restriction of Fourier transform to quadratic surfaces and
decay of solutions of wave equations, Duke Math. J. {\bf 44} (1977), 705-774.

\bibitem{sw} A. Soffer and M.I. Weinstein, Multichannel nonlinear scattering for nonintegrable
equations II. The case of anisotropic potentials and data, J.  Differential Equations 
{\bf 98} (1992), 376-390.  
\bibitem{we1} R. Weder, Inverse scattering for the nonlinear Schr\"odinger equation,
Comm. Partial Differential Equations {\bf 22} (1997), 2089-2103. 

\bibitem{we2}R. Weder, $L^p-L^{\acute{p}}$ estimates for the Schr\"{o}dinger equation on the line
and inverse scattering for the nonlinear Schr\"{o}dinger equation with a potential, J. Funct.
Anal. {\bf 170} (2000), 37-68.

\bibitem{we5} R. Weder, Center manifold for nonintegrable nonlinear Schr\"odinger equations
 on the line, Comm. Math. Phys. {\bf 215} (2000), 343-356.

\bibitem{we3} R. Weder, Inverse scattering for the non-linear Schr\"odinger equation: 
reconstruction of the potential and the nonlinearity, Math. Methods  Appl.
Sci. {\bf 24} (2001), 245-254. 

\bibitem{we4} R. Weder, Inverse scattering for the nonlinear Schr\"odinger equation II.
Reconstruction of the potential and the nonlinearity in the multidimensional case,
Proc. Amer. Math. Soc. {\bf 129} (2001), 3637-3645.

\bibitem{we6} R. Weder, Scattering for the forced non-linear Schr\"odinger equation with a potential on the half-line,
preprint, 2002.
\bibitem{wei} J. Weidmann, "Spectral Theory of Ordinary Differential Operators", Lecture Notes
in Math. {\bf 1258}, Springer, Berlin, 1987.

\bibitem{ya1} K. Yajima, The $W^{k,p}$-continuity of wave operators for Schr\"{o}dinger operators,
J. Math. Soc. Japan {\bf 47} (1995), 551-581.

\bibitem{ya2} K. Yajima, The $W^{k,p}$-continuity of the wave operators for Schr\"odinger 
operators. III. Even dimensional cases $m \geq 4$, J. Math. Sci. Univ. Tokyo {\bf 2} (1995),
311-346.     

\bibitem{ya3} K. Yajima, $L^p$-boundedness of wave operators for two dimensional Schr\"{o}dinger
operators, Commun. Math. Phys. {\bf 208} (1999), 125-152.

  

 \end{thebibliography}
\end{document}